\documentclass{elsart}
\usepackage{graphicx}
\usepackage{amsmath,amssymb}
\usepackage[english]{babel}

\newcommand{\sted}{\mathrm{S}}
\newcommand{\inflow}{\mathrm{bnd}}
\newcommand{\hpf}{\mathrm{H}}
\newcommand{\trn}{\mathrm{T}}
\newcommand{\fdscrit}{\mathrm{FDS}}
\newcommand{\fdssol}{\mathrm{FDS}}
\newcommand{\cnab}{\mathrm{ca}}

\newcommand{\difadv}{r}
\newcommand{\linstab}{0}
\newcommand{\dificrit}{\mathrm{DIFI}}

\newcommand{\onefig}[1]{\includegraphics[width=0.8\columnwidth]{#1}}
\newcommand{\panelfig}[2]{\raisebox{10pt}{#1)}\onefig{#2}}

\begin{document}

\begin{frontmatter}

\title{Flow- and Diffusion Distributed Structures with noise at the inlet}

\author[SGAP]{Pavel V. Kuptsov\corauthref{cor}}\ead{p.kuptsov@rambler.ru},
\author[CITY]{Razvan A. Satnoianu}

\address[SGAP]{Department of
Informatics, Saratov State Law Academy, Volskaya 1, Saratov
410056, Russia}
\address[CITY]{Faculty of Actuarial Science, Cass Business School,
City University, London EC1Y 8TZ, UK}

\corauth[cor]{Corresponding author.}

\begin{abstract}
Flow and Diffusion Distributed Structures (FDS) are stationary
spatially periodic patterns that can be observed in
reaction-diffusion-advection systems. These structures arise when
the flow rate exceeds a certain bifurcation point provided that
concentrations of interacting species at the inlet differ from
steady state values and the concentrations are held constant.
Normally, theoretical studies of these patterns are developed
without concerning a noise. In this paper we consider FDS for a
more realistic conditions and assume that the inlet concentrations
are perturbed by a small noise. When the flow rate is small, the
FDS is linearly sensitive to noise at the inlet. Even weak
fluctuations destroy the stationary pattern and an oscillatory
solution appears instead. For higher flow rates the instability
becomes nonlinear: the pattern remains unaltered for a weak noise
and undergoes the destruction when the noise amplitude passes a
certain threshold. We represent a detailed description of these
effects and examine two scenarios for the stabilization.
\end{abstract}


\end{frontmatter}

\section{Introduction}
Reaction-diffusion system are known to demonstrate a variety of
instabilities that result in stationary patterns or oscillatory
behavior. Flow and Diffusion Distributed Structure (FDS) is
relatively new type of patterns that can appear in a reaction
diffusion system in presence of an open flow. The necessary
condition for this pattern to appear is oscillatory instability
and constant, non steady state concentrations of interaction
species at the inlet. If these condition are satisfied, the FDS
appears when the flow rate passes above a certain critical point.
Patterns of this type were first described by Kuznetsov et
al.~\cite{KuzMos97} and later they were studied in more details by
Andres{\'e}n et al.~\cite{AndrBache99}. Soon after discovery these
patterns were observed in the experiments by K{\ae}rn and
Menzinger~\cite{KernMenz99,KernMenz00b}. Later the experimental
results were summarized in a survey paper~\cite{KernMenz02}.

There is a bit confusing variety of terms concerning this type of
pattern formation phenomena. Initially, K{\ae}rn and Menzinger
suggested the term ``Flow distributed oscillations''
(FDO)~\cite{KernMenz99}. Later, the abbreviation FDS was suggested
by Satnoianu~\cite{SatMaini01,Sat03}. In this paper we adhere to
the last version and refer to the structures under consideration
as FDS. (See also the discussion of the terminology
in~\cite{McGrawMenz05a}).

Extensive theoretical studies of new structures are provided
in~\cite{SatMenz00,SatMaini01,Sat03,McGrawMenz05a}. In particular,
these papers explore the relationship between FDS and other known
types of instabilities in reaction-diffusion-advection systems:
Turing patterns, DIFI and Hopf instabilities. As shown by McGraw
and Menzinger in~\cite{McGrawMenz05a}, the FDS (they use the term
FDO, see the remark above) is closely related to the Hopf
instability and DIFI, while the Turing pattern has a different
nature.

As mentioned before, an FDS pattern emerges provided that a system
demonstrates an oscillatory instability. Often a convective
instability of the oscillatory mode is declared as a necessary
condition for FDS: arbitrary small constant perturbation to the
steady state at the inlet can grow to an FDS. But as shown by
Kuptsov~\cite{Kup04}, the FDS can also appear in an absolute
instability domain. In this case, however, the transition to FDS
is rigid. If the inlet perturbation is small, the oscillatory
solution dominates an FDS. However, the increasing of the
perturbation gives an advantage to the FDS and it grows and
suppresses the oscillations.

An experimental study of an FDS in presence of a differential
transport is provided by M\'{\i}guez et
al.~\cite{MigSat06,MigIzus06}. They consider a remarkable variety
of impacts on an FDS pattern. These are a parallel formation of
two neighboring FDSs, an FDS in a system with monotonically
increasing flow rate, an FDS in 2D system with sinusoidally
varying $y$-boundary and also interaction of two perpendicular
FDSs. This analysis reveals a high degree of robustness and
structural stability of this type of patterns.

A study of forcing to FDS is also presented in a work by Kuptsov
et al. in~\cite{KupKuz02}. This paper provides a numerical
analysis of an FDS that is perturbed by a small particle dragged
by a flow. Papers~\cite{KernMenz00b,BamToth02,McGrawMenz05b} are
devoted to the study of an FDS in presence of periodic
oscillations at the inlet. We shell discuss these papers below in
the final section, because they are closely related to our
analysis.

Recently, Kuptsov at al. represented a theoretical analysis of
FDSs in a 2D system with Poiseuille flow~\cite{KupSatDan05}. Two
basic types of patterns were observed. The first one is
$y$-dependant version of 1D FDS. The pattern consists of curved
stationary stripes that are transverse to the flow direction.
Patterns of this type were also observed in the experiments by
M\'{\i}guez et al.~\cite{MigIzus06}. The second type of structures
is specific to the 2D case: the structure consists of several
stationary longitudinal stripes. If both types of basic structures
are allowed, a more complicated pattern appears as a combination
of transverse and longitudinal stripes.

All previous theoretical approaches have been developed for
perfectly constant condition at the inlet. But fluctuations, that
are always present in natural systems, can influence the pattern
formation. As observed in many experiments, see for
example~\cite{KernMenz02} and~\cite{TayBam02}, FDS near the
critical point can be suppressed by oscillatory solutions induced
by the inlet fluctuations.

The effect of an inlet noise on convectively unstable distributed
systems was studied by many authors. We recall that in the
convectively unstable system a growing perturbation drifts
downflow but decays when observed from any fixed point. By
contrast, absolute instability implies that the perturbation,
being initially localized, grows at any fixed point in
space~\cite{Saar88,Saar89,KuzMos97,KupKuz03} (see also reviews
therein). In view of the extreme sensitivity of the convectively
unstable state to perturbations, Deissler~\cite{Deissler85},
Deissler and Farmer~\cite{DeissFarm92}, and Borckmans et
al.~\cite{BorckDew95} conclude that wave-patterns may be generated
and sustained due to amplification of fluctuations from an
upstream noise source. More subtle than a straightforward
amplification of noise, an interplay between fluctuations and
dynamics is suggested by Landa~\cite{Landa96} who treat the onset
of turbulence in flow systems as a kind of noise-induced
transition.

Kuznetsov~\cite{Kuz02} studied the effect of inlet noise on a
spatially uniform reaction-diffusion-advection system. (This
problem is similar to the ours, but we also apply a constant
perturbation at the inlet to obtain an FDS pattern.) A
noise-induced absolute instability was observed: the inlet noise
increases the critical flow rate for transition from absolute to
convective instability, and the new critical value is proportional
to the square root of the noise amplitude, i.e., it obeys a power
law.

In our previous paper~\cite{KupSat05} we already discussed the
stability of FDSs to an inlet noise. In particular, we observed
that when the flow rate is close to the FDS critical point, the
noise, regardless of its amplitude, destroys the structure, i.e.,
the FDS is linearly unstable. For higher flow rates the system
passes a new critical point, thus becoming insensitive to a weak
noise. Beyond this point, noise destroys the structure only when
the noise amplitude exceeds a threshold value. The threshold grows
as a power of the flow rate. We shall refer to this case as
nonlinear instability of an FDS.

The purpose of the present paper is to extend our previous study.
We demonstrate two scenarios of stabilization and develop a linear
stability analysis of an FDS pattern. Sec.~\ref{sInstab} reviews a
variety of transitions to FDS. Sec.~\ref{sStab} represents a
qualitative picture of the destruction and stabilization of FDSs
by a noise. In Sec.~\ref{sTwoScen} we summarize previous
considerations and, employing Fourier analysis, introduce two
scenarios of the stabilization. In Sec.~\ref{sLin} this two
scenarios are confirmed by a linear stability analysis. Finally,
Sec.~\ref{sConcl} summarizes our results.

\section{\label{sInstab}Oscillatory instabilities and transition to FDS}

Before starting the main analysis let us introduce our model
system and briefly review the oscillatory instabilities that can
produce FDS patterns.

We shall consider the well-known Lengyel-Epstein model for the
CIMA (chlorite-iodide-malonic acid-starch)
reaction~\cite{LengEpst91,JenPanMos94,SatMaini01}, assuming that
the reagents are carried by a flow in a one-dimensional reactor:
\begin{subequations}
  \label{eLengEpst}
  \begin{eqnarray}
    u_t + \phi u_x - u_{xx}  &=& a - u - 4 u v / \left(1+u^2\right),\\
    v_t + \difadv \phi v_x - \delta v_{xx} &=& b \left( u - u v / \left(1 + u^2\right) \right).
  \end{eqnarray}
\end{subequations}
Here $u$ and $v$ denote the dynamical variables related to the
concentrations of two chemical species, $a$ and $b$ are treated as
control parameters, and $\phi$ is the flow rate. The transport
rates for $u$ and $v$ can be different: $\delta$ introduces the
differential diffusion, and $\difadv$ controls the differential
flow. We shall study two cases: no differential flow at
$\difadv=1$ and identical ratios for diffusion and flow at
$\difadv=\delta$. The reactor is supposed to be semi-infinite.

Note that we use a nonstandard form of Eqs~(\ref{eLengEpst}).
Normally, in the model equations for CIMA reaction parameters
$\delta$ and $b$ are proportional to each other. But in all our
studies we always take \emph{a constant} values for the parameters
and analyze the varying of the flow rate or noise amplitude. So,
at each step of observations one can easily recalculate our
parameters to the standard ones. But we believe that the specific
numerical values are not very important because we do not study
the system near bifurcation points. As we have examined, the
represented results remain valid for a wide variety of the
parameter values.

We shall find numerical solutions to Eqs.~(\ref{eLengEpst}) via
the semi-implicit Cranck-Nicholson method with steps of
discretization $\Delta x \approx 0.1$ and $\Delta t \approx
0.0125$. Also, in all our numerical simulations an outlet
condition $(u_x,v_x)_{x=X}=0$ is applied, where $X$ is the length
of the reactor. This type of the outlet conditions are normally
used in numerical simulations of reaction-diffusion systems, see
for example~\cite{JenPanMos94,KuzMos97}. These conditions are the
best choice for a model of a semi-infinite, i.e., very long,
system. In such a system the right boundary should not influence
an overall picture. The vanish of $u_x$ and $v_x$ at the outlet
effectively prevents an upstream propagation of any numerical
artifacts. They, if appear at all, stay near the right boundary as
a tiny area of oscillations.

The system~(\ref{eLengEpst}) admits the homogeneous steady state
solution
\begin{equation}
  \label{eHSS}
  u_\sted = a/5, \;\; v_\sted = 1 + a^2/5^2.
\end{equation}
In the absence of a flow the steady state can be destabilized
either by Hopf or by Turing modes whose growth rates becomes
positive at
\begin{equation}
  b=b_\hpf = (3a^2 - 125)/(5a)
\end{equation}
and
\begin{equation}
  b=b_\trn = (125+13a^2-4a\sqrt{10(25+a^2)})\delta)/(5a),
\end{equation}
respectively~\cite{Turing52,NicPrig77,JenPanMos94}. The Hopf
instability gives rise to uniform temporal oscillations while the
Turing instability produces a stationary space-periodic pattern.
Note that for the considered system supercritical values of $b$
lie below the bifurcation points.

When both components of the system are transported by a flow with
identical rates, i.e., $\difadv=1$, the Hopf instability remains
unaltered while the Turing structure loses its stationarity. (Of
course, this takes place only for an observer in a laboratory
reference frame. The other one that moves with the flow still
registers normal Turing stripes). The frequency of temporal
oscillations of the appearing waves is proportional to the flow
rate~\cite{KernMenzSat02}.

Differential flow with $\difadv\neq1$ results in the formation of
new spatio-temporal structures due to the so-called
Differential-Flow Instability (DIFI) \cite{RovMenz93}. The most
interesting situation is observed outside the Hopf and Turing
domains. The homogeneous steady state of the system remains stable
until the flow rate reaches a critical point $\phi_\dificrit$.
Above this point, a DIFI mode become unstable and give rise to
travelling waves.

In presence of a flow an instability can be either absolute or
convective. The absolute instability appears for small flow rates
and manifests itself as a spreading of an initially localized
perturbation over the whole space. For higher flow rates the
instability becomes convective: the growing perturbation, being
carried by the flow, decays in any fixed point. A critical point
$\phi_\cnab$ for the transition from absolute to convective
instability can be obtained from dispersion equation of a system
as described in~\cite{Saar88,Saar89,KuzMos97,KupKuz03} (see also
reviews therein).

\begin{figure}
  \centering
  \panelfig{a}{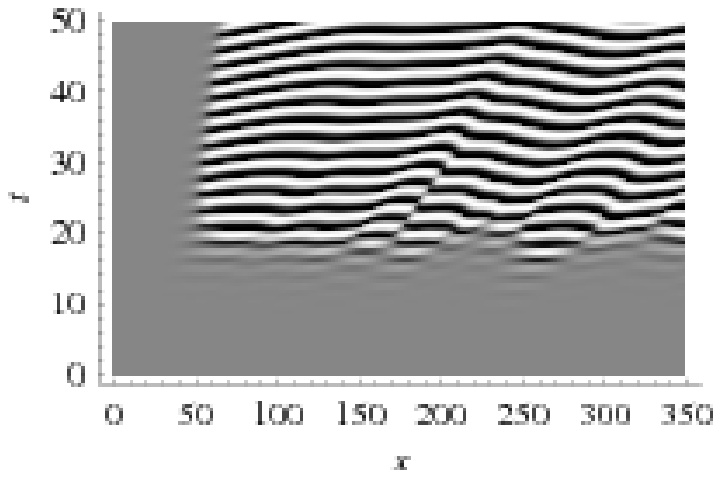}\\
  \panelfig{b}{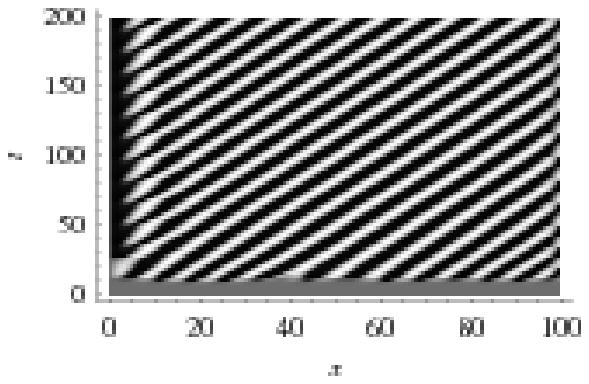}\\
  \panelfig{c}{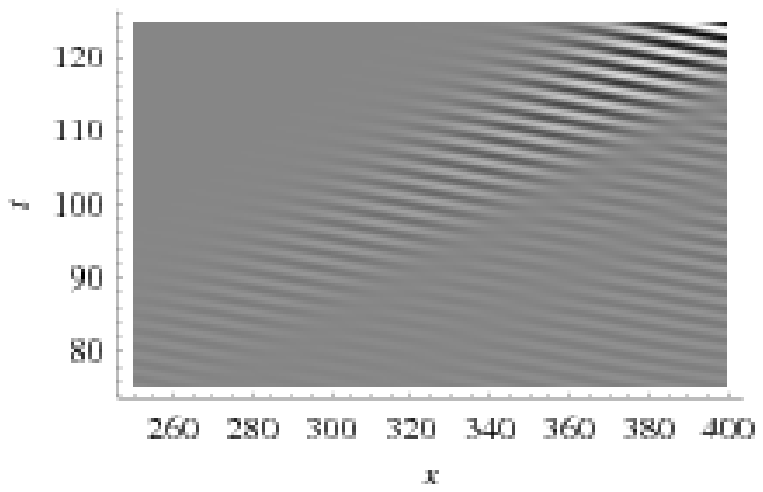}
  \caption{\label{fInstab}Oscillatory instabilities in the system~(\ref{eLengEpst}).
  Grey levels indicate values of $u$. For all panels $a=20$.
  (a) Hopf instability at $b=6$, $\delta=3$, $\difadv=1$, $\phi=3$.
  (b) Turing instability at $b=6$, $\delta=15$, $\difadv=1$, $\phi=0.5$.
  (c) DIFI at $b=12$, $\delta=\difadv=3$, $\phi=1.5$.}
\end{figure}

Figure~\ref{fInstab} represents solutions to
Eqs.~(\ref{eLengEpst}) that are induced by the instabilities
described above. In the other words, this figure demonstrates what
happens in our system without an FDS. For all panes of the figure
the initial condition at $t=0$ is the homogeneous steady state
perturbed by small, spatially inhomogeneous fluctuations. The
inlet is held at the steady state.

Figure~\ref{fInstab}(a) illustrates a convective Hopf instability.
We observe a temporally periodic structure with a wave number that
is zero on average. Because the initial distribution of the
reacting species is inhomogeneous, Hopf oscillations appears in
different points with different initial phases. Zero wave number
of a Hopf mode means that the effective interaction between
neighboring points is very weak. As a result we observe in the
figure a lot of local phase perturbations. The perturbations,
however, are slowly smoothed out by the diffusion. These
perturbations as well as the upstream edge of the structure travel
downflow. This confirms the convective nature of the observed
instability. Note that this is dictated by a dispersion equation
and does not depend on boundary or initial conditions. One can
easily reproduce this convective moving of the spatio-temporal
structure for different conditions at the system edges or for
another inial perturbation.

Figure~\ref{fInstab}(b) represents a spatio-temporal structure
that appears due to an absolute Turing instability. The initial
state for this figure is inhomogeneous, but, in contrast to the
Hopf case, the structure forms due to the spatial interaction
stimulated by a differential diffusion. This process subordinates
the concentrations of the reacting species in different points and
results in the perfect spatially periodic pattern. In a
spatio-temporal diagram this should looks like a vertical stripes.
But a flow transports the Turing structure. As a result we observe
that the stripes are sloped exactly along the flow. (In the
spatio-temporal diagram $\phi=0.5$, $t_\mathrm{max}=200$ and
$x_\mathrm{max}=\phi t_\mathrm{max}=100$, so that the flow is
parallel to the main diagonal).

Finally, Fig.~\ref{fInstab}(c) illustrates a DIFI structure. For
the parameters values that are used for this figure, the system is
stable without a flow, but different flow rates of the reactants
produces an instability. Now both frequency and wave number are
nonzero. Observe that they have opposite signs, and DIFI waves
travel upstream. The DIFI takes place at $\phi_\dificrit$ that is
normally higher than the critical value for absolute-convective
instability transition $\phi_\cnab$. In our case the DIFI is also
convective, so that the structure is washed by a flow.

FDS patterns appear in the system when temporal oscillations are
spread in space by a flow. Every volume of an oscillating medium,
being carried by the flow, appears with different phases in
different spatial points and thereby produces a spatially periodic
structure. The flow rate is constant, so this structure is
stationary if all portions have identical initial phases. For this
to be fulfilled, a constant boundary perturbation should be
applied at the inlet. This is a permanent displacement of reagent
concentrations from the homogeneous steady state. A wave number of
FDS can be roughly estimated as $\omega / \phi$, where $\omega$ is
the frequency of the oscillations registered in a reference frame
that moves with the flow (see discussion
in~\cite{AndrBache99,KernMenz99,AndrMos00,KernMenz00}).

A Hopf mode is responsible for the formation of FDS when all
components of the system have identical flow rates. In the
presence of a differential flow this role is played by a DIFI
mode. Turing oscillations can not produce an FDS because this
solution is stationary in the co-moving reference frame, so that
for a stationary observer every portion of a medium moves along
the system with constant phase. This simple conclusion is in a
good agreement with the analysis provided by McGraw and
Menzinger~\cite{McGrawMenz05a}: the FDO/FDS patterns are shown to
be closely related to Hopf instability and DIFI, while the Turing
pattern formation mechanism has a quite different nature.

To realize the formation of FDS, it is enough to perturb only one
of the components and keep steady state value for the other:
\begin{equation}
  \label{eConstInlet}
  u(x=0,t) = u_\sted + u_\inflow, \;\; v(x=0,t) = v_\sted,
\end{equation}
where $u_\inflow$ is a constant perturbation of the boundary
value. As before, $u_\sted$ and $v_\sted$ represent the
homogeneous steady state. When the flow rate is small, the
perturbation decays in space. A non-decaying FDS solution appears
above a critical point~\cite{KuzMos97,AndrBache99,SatMaini01},
\begin{equation}\label{eFDSCrit}
  \phi_\fdscrit = \sqrt{\frac
    {40 a^3 b {(\delta + \difadv)}^2 - {(3 a^2 \delta + 5 a b - 125 \delta)}^2 \difadv}
    {(25 + a^2 ) (\delta  + \difadv)(3 a^2 \difadv - 5 a b - 125 \difadv) \difadv}
  }.
\end{equation}
This critical value diverges at $b=\difadv b_\hpf$. Without a
differential flow, i.e., at $\difadv=1$, the condition of
divergence coincides with the condition for the Hopf instability,
while at $\difadv>1$ an FDS appears even if the flow-less system
is stable.

If $\phi>\phi_\fdscrit$ and $\phi>\phi_\cnab$, the transition to
FDS takes place when an oscillatory mode is convectively unstable.
Because in this case growing oscillations travel with the flow,
any small inlet perturbation can freely grow to the FDS pattern.
This case is referred to as a soft transition to
FDS~\cite{KupKuz03,Kup04}. Otherwise, if $\phi_\fdscrit < \phi <
\phi_\cnab$, the oscillatory mode is absolutely unstable and tends
to occupy the whole space. Now the FDS can grow only when the
inlet perturbation is high, because it must overcome the
competition from the oscillatory solution. This is a rigid
transition to FDS~\cite{KupKuz03,Kup04}. In this paper we analyze
FDS for $\phi>\phi_\cnab$, i.e., we are interested in the case of
the soft transition.

\section{\label{sStab}FDS in presence of inlet noise: a qualitative picture}

We introduce a noise via the following boundary condition at the
inlet:
\begin{equation}
  u(x=0,t) = u_\sted + u_\inflow (1 + \theta \: \xi(t)), \; v(x=0,t) = v_\sted.
\end{equation}
Here, $u_\sted$ and $v_\sted$ denote the homogeneous steady state
and $u_\inflow$ is the constant boundary perturbation. $\xi(t)$
describes a noise with the uniform distribution in the interval
$[-1, 1]$, and $\theta$ controls an amplitude of the noise
relating to $u_\inflow$. The particular distribution of a noise is
not very important. But, as we explain below, it should be
sufficiently wide to contain required harmonics.

In our previous work~\cite{KupSat05} we have studied the stability
of FDS patterns in the Hopf oscillatory domain. In this paper we
consider the following four cases.
\begin{enumerate}
\item FDS appears in presence of the Hopf instability: $\difadv=1$ and
$b_\trn<b<b_\hpf$.

\item FDS is observed in presence of a differential flow,
$\difadv>1$. The system is stable without a flow, i.e., $b>b_\hpf$
and $b>b_\trn$. Oscillations appear due to DIFI when the flow rate
is above a critical point $\phi_\dificrit$.

\item The transition to FDS takes place when both Turing and
Hopf modes are unstable: $\difadv=1$ and $b\ll b_\hpf<b_\trn$. The
Turing mode has a higher growth rate, but the Hopf mode is also
significant.

\item FDS appears again in the presence of Hopf and Turing modes, but
the system is close to the Hopf point: $b\lesssim b_\hpf$. Growth
rate of the Hopf mode is small, so that this mode is much weaker
than the Turing one.
\end{enumerate}

The analysis developed in~\cite{KupSat05} shows that in presence
of an inlet noise the FDS is destroyed at small flow rates and
appears only when the flow rate passes a certain point of
stabilization. The stabilization flow rate is found to depend on
the noise amplitude as a power law. For a fixed flow rate the FDS
is stable if the noise amplitude is small and the pattern is
destabilized and destroyed when the amplitude passes a certain
threshold. In the present paper we refer to this phenomenon as a
nonlinear instability of FDS. The picture differs for flow rates a
bit above the FDS critical point $\phi_\fdscrit$. The threshold
noise level absent and FDSs are linearly unstable so that any
small inlet fluctuations destroy the pattern. We already reported
this case in~\cite{KupSat05}. In Sec.~\ref{sLin} a detailed
investigation is provided.

\subsection{Case 1: Hopf instability}

\begin{figure}
  \centering
  \panelfig{a}{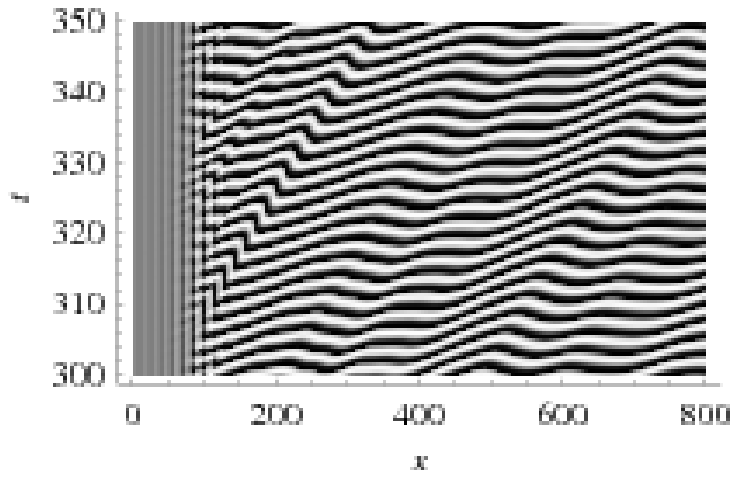}\\
  \panelfig{b}{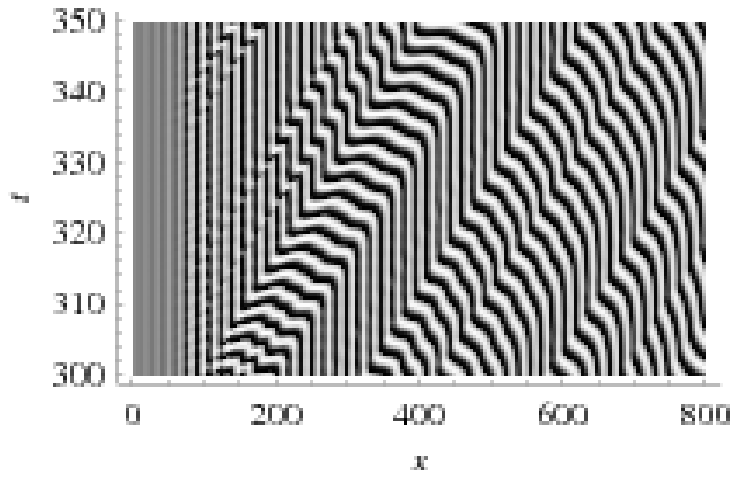}
  \caption{\label{fSptmHopf}FDS in the presence of inlet noise when
  a Hopf mode is unstable.
  In panel (a) the noise destroys the FDS pattern, while in panel (c)
  (see below) the FDS is stabilized at a higher flow rate.
  Flow rate grows from (a) $\phi=6$ to (b) $\phi=6.5$ and (c) $\phi=7$.
  $a=20$, $b=6$, $\delta=3$, $\difadv=1$, $u_\inflow=0.1$, $\theta=0.05$
  ($b_\hpf=10.75$, $b_\trn\approx 3.289$).}
\end{figure}

\begin{figure}
  \addtocounter{figure}{-1}
  \centering
  \panelfig{c}{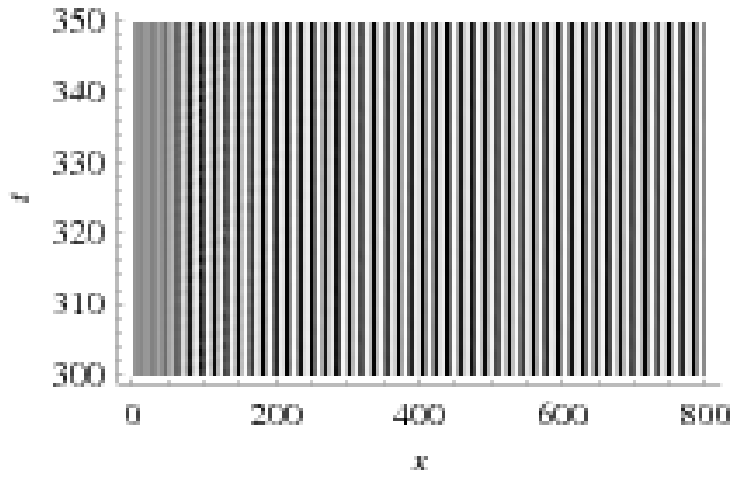}
  \caption{(\emph{continued}) }
\end{figure}

Figure~\ref{fSptmHopf} demonstrates an interaction between an FDS
and inlet noise when only a Hopf mode is unstable. The constant
part of the inlet perturbation is small, so that the FDS in this
figure has a boundary layer. A structure without a boundary layer
is stabilized at smaller flow rate and has a number of
peculiarities that we shall discuss below.

In Fig.~\ref{fSptmHopf}(a) the FDS pattern disappears after
several periods and a new oscillatory structure forms instead.
This effect can be explained by a selective amplification of
Fourier components of the noise. The maximum of this amplification
corresponds to a Hopf mode~\cite{KupSat05}. This results in a
resonant excitation of this mode by the noise so that a growing
oscillatory solution destroys the FDS. A corresponding Hopf
structure without a noise is shown in Fig.~\ref{fInstab}(a). Note
a high degree of similarity between noise and noiseless solutions.

If an FDS has a boundary layer, i.e., $u_\inflow$ is small, the
amplification basically occurs inside this layer. A perturbation
to the pattern always manifests itself at the end of the boundary
layer regardless of the flow rate, compare
Figs.~\ref{fSptmHopf}(a) and (b). If $u_\inflow$ is high and the
boundary layer is absent, the noise also destroys an FDS pattern.
But an effective amplification ratio of a fully developed FDS
pattern is much lower, so that the length of an unperturbed FDS
area can be very long. (We do not provide an illustration because,
except for the boundary layer, the spatio-temporal diagram looks
as that in Fig.~\ref{fSptmHopf}(a)).

Figure~\ref{fSptmHopf}(b) illustrates an intermediate situation
for a higher flow rate. Now the influence of the noise is weaker
and the FDS is not totally destroyed. The overall structure is a
composition of vertical fragments of FDS and sloped stripes of an
oscillatory solution. Observe a remarkable regularity of this
composition. This interesting structure appears because both the
stationary stripes and the oscillatory fragments are different
manifestations of a Hopf solution. (Recall that FDS is, in fact,
an oscillations that are spread in space due to a flow and
diffusion.) This explains why the fragment are joined so well into
continuous step-like stripes.

This type of structures is a typical response of an FDS to a
perturbation. K{\ae}rn and Menzinger~\cite{KernMenz00b} first
observed them in a system with oscillatory inlet
boundary~\cite{KernMenz00b}. The structures was referred to as
pulsating waves. A more complicated zigzag patterns are reported
by Taylor et al.~\cite{TayBam02}. Also, similar structures was
described by Kuptsov et al.~\cite{KupKuz02} for an FDS perturbed
by a moving particle. In addition, coexisting stationary and
travelling waves have been observed in a system without
noise~\cite{Sat03}.

\begin{figure}
  \centering
  \onefig{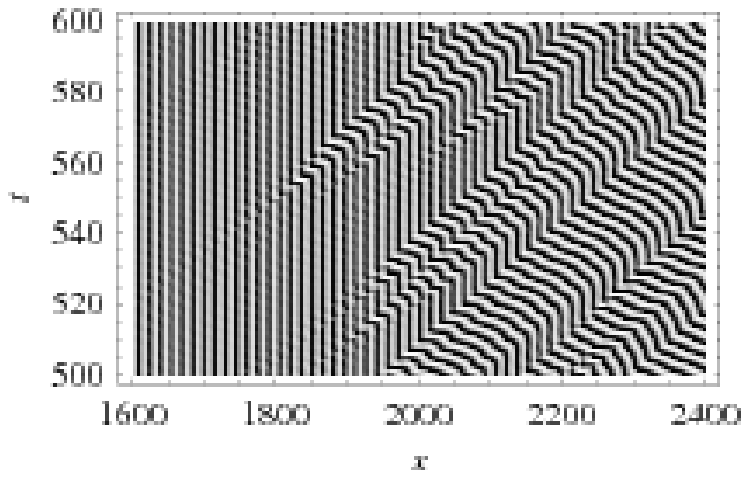}
  \caption{\label{fNoLayer}Spatio-temporal intermittency for an
  FDS without a boundary layer. The parameters are as in Fig.~\ref{fSptmHopf}
  except $\phi=6.1$ and $u_\inflow=2$. Unlike
  Fig.~\ref{fSptmHopf}(b), the structure appears far from the inlet.}
\end{figure}

FDS without a boundary layer can also demonstrate spatio-temporal
structures composed of stationary and oscillatory fragments. An
example is shown in Fig.~\ref{fNoLayer}. As mentioned above, for
FDS without a boundary layer the amplification area can be very
extended. In Fig.~\ref{fNoLayer} the FDS at $x<1800$ looks like a
noiseless pattern. However, beyond this point noise demonstrates
itself. Note that the flow rate in Fig.~\ref{fNoLayer} is less
then in Fig.~\ref{fSptmHopf}(b). In the other words, an FDS
without a boundary layer is less sensible to noise and becomes
stable at a smaller value of the flow rate.

Further increase of the flow rate results in the growth of the
stationary fragments and shrinking of the oscillating ones.
Finally, all oscillating fragments disappear and we observe a
pattern as in Fig.~\ref{fSptmHopf}(c). The structure here looks
like a perfect FDS without a noise.

\begin{figure}
  \centering
  \onefig{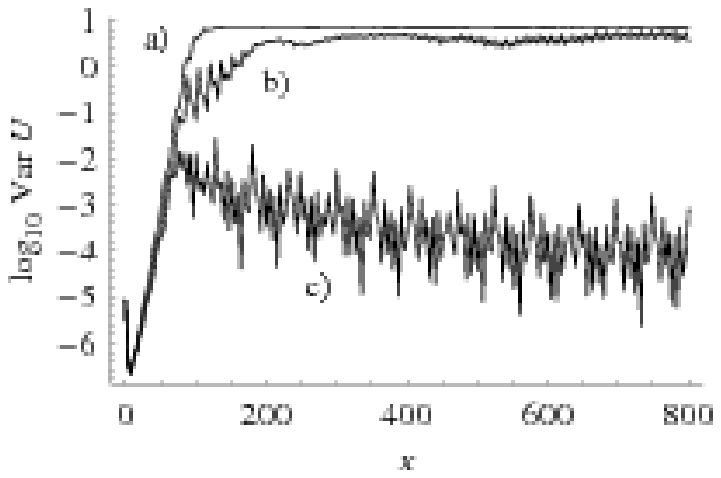}
  \caption{\label{fVar}Spatial distributions of the variances of temporal
  oscillations before, curves (a) and (b),
  and after the stabilization of FDS, curve (c).
  The curves correspond to the panels in Fig.~\ref{fSptmHopf}
  with respective letters.}
\end{figure}

Figure~\ref{fVar} provides a more accurate verification of the
stabilization. This figure shows spatial distributions of
variances of temporal oscillations. The distributions are
calculated for the diagrams in Fig.~\ref{fSptmHopf}. Labels (a),
(b) and (c) on the curves correspond to the panels of
Fig.~\ref{fSptmHopf}. All curves grow exponentially within the
boundary layer of the FDS and nearly coincide there. In the other
words, the boundary layer amplifies noise even when the
corresponding fully developed FDS is stable. Behind the boundary
layer the curves behave different. Curve (a) continues to grow and
reaches a saturation very fast. This corresponds to the
oscillations observed in the right part of
Fig.~\ref{fSptmHopf}(a). The intermediate curve (b) also grows,
but does not tend to saturation. It varies slowly and irregularly
near a relatively high value. This occurs because time series in
this case are composed of fragments of two solutions, see
Fig.~\ref{fSptmHopf}(b). They are irregularly arranged and their
sizes are distributed within a wide range. Curve (c) confirms the
stabilization. The variance decays, therefore, the FDS remains
stable even far downflow.

\begin{figure}
  \centering
  \onefig{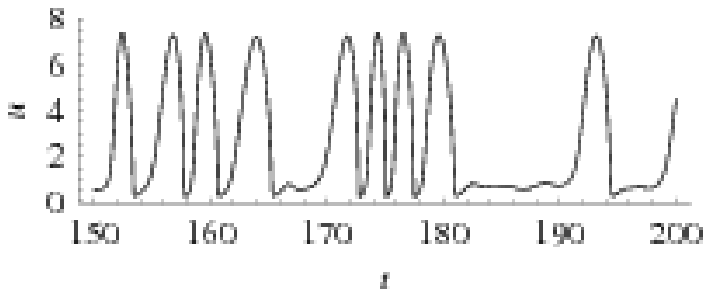}
  \caption{\label{fIntermit}Intermittent time series of $u$
  recorded in a fixed spatial point for the case represented in
  Fig.~\ref{fSptmHopf}(b). Laminar phases correspond to fragments
  of FDS stripes.}
\end{figure}

The combination of stationary and oscillatory solutions that is
demonstrated in Figs.~\ref{fSptmHopf}(b) and \ref{fNoLayer} can be
treated as a spatio-temporal intermittency. A similar compound
structure has been analyzed by Kuptsov et al.~\cite{KupKuz02}.
Figure~\ref{fIntermit} demonstrates an example of an intermittent
time series. Variable of $u$ is observed from a fixed point in
space: laminar phases, that correspond to fragments of FDS, are
interrupted by bursts of oscillations.

\begin{figure}
  \centering
  \onefig{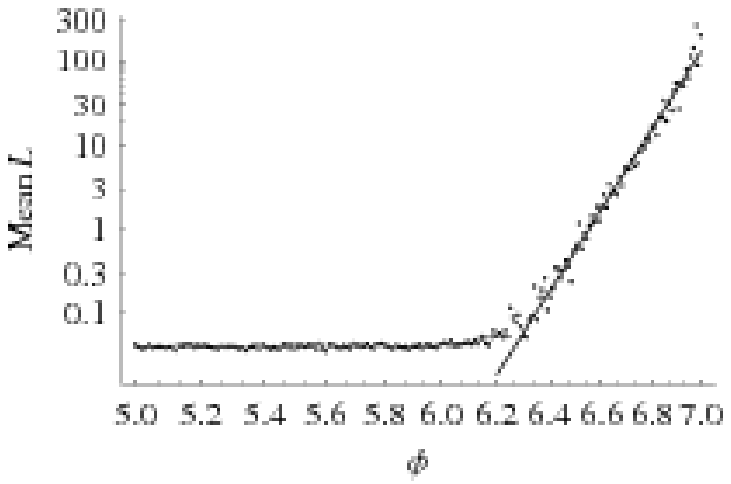}
  \caption{\label{fLaminar}Mean lengthes of FDS fragments vs. flow rate measured in a fixed spatial point.
  The parameters are as in Fig.~\ref{fSptmHopf}.
  The dots, computed numerically, are plotted in double logarithmic scale. In the right part
  the numerical data are approximated well by a straight line that
  indicates a power-law dependence. The exponent, equal to the slope of the line,
  is $\gamma\approx 73$.}
\end{figure}

A characteristic feature of intermittency is a power-law
divergence of the mean length of laminar phases as a bifurcation
parameter approaches a critical point. In our case the flow rate
controls the mean length of FDS fragments and the divergence
corresponds to the stabilization of FDS. Let us employ the
following algorithm: Given the flow rate, we compute a pure FDS
pattern without a noise. Then we take a point at some distance
from the inlet where the FDS is definitely fully developed (for
all computations we take the same point). Starting from this
position, within the period of the FDS we seek another point where
$u$ is in the middle between its maximum and minimum values. The
value of $u$ in this point is treated as a laminar state. Now a
noise is switched on and after a transient period the mean time of
staying near the laminar state is registered. Repeating this
procedure for different flow rates, we obtain the dependence shown
in Fig.~\ref{fLaminar}. When the flow rate is small, $u$
oscillates, as shown in Fig.~\ref{fSptmHopf}(a), and passes the
laminar state very fast. In Fig.~\ref{fLaminar} the corresponding
mean values are very small. Intermittency appears for $\phi>6.2$,
and the mean time spent near the laminar state starts to grow with
the flow rate. The numerical data in this part of the figure are
approximated well by a straight line in double logarithmic scale.
This implies a power-law dependence
$<L>=(\phi-\phi_\mathrm{c})^\gamma$. Note that the exponent
$\gamma\approx 73$ is extremely high.

\subsection{Case 2: DIFI}

\begin{figure}
  \centering
  \panelfig{a}{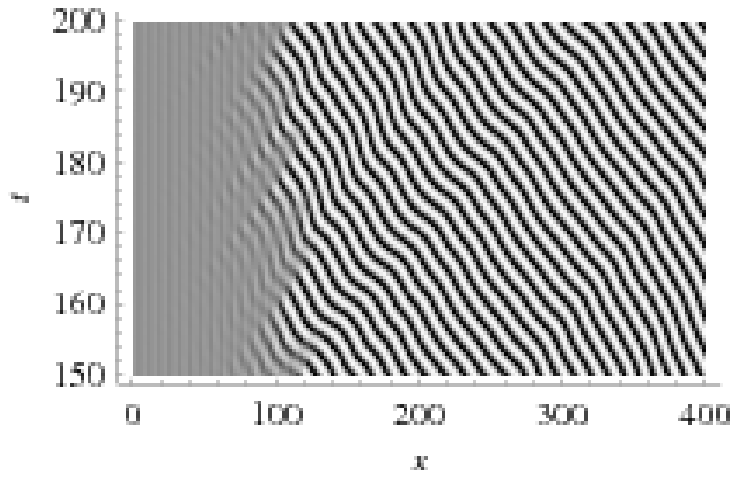}\\
  \panelfig{b}{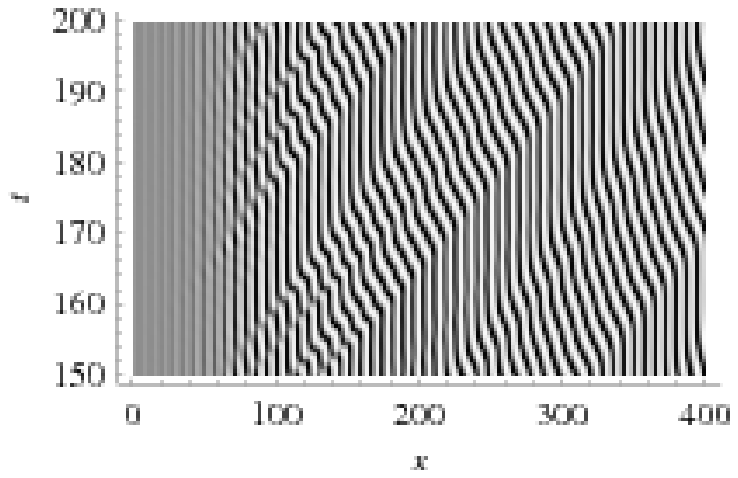}
  \caption{\label{fSptmDIFI}FDS with noise when a DIFI mode is unstable:
  (a) $\phi=2.3$, and (b) $\phi=2.47$.
  $a=20$, $b=12$, $\delta=\difadv=3$, $u_\inflow=0.1$, $\theta=1$
  ($b_\hpf=10.75$, $b_\trn\approx 3.289$, $\phi_\dificrit\approx 1.47$ and
  $\phi_\fdscrit\approx 2.27$).}
\end{figure}

This case, illustrated in Fig.~\ref{fSptmDIFI}, is similar to the
previous one. In Fig.~\ref{fSptmDIFI}(a) we observe how a noise
excites an unstable oscillatory mode that grows to an oscillatory
solution. But now a DIFI mode is responsible for the formation of
FDS. Hence, it is the DIFI solution that appears behind the
destructed FDS. Observe the high regularity of the appearing
structure. In Fig.~\ref{fSptmDIFI}(b) a higher flow rate results
in the formation of spatio-temporal intermittency similar to the
structure in Fig.~\ref{fSptmHopf}(b). A perfect correspondence
between stationary and oscillatory fragments is also observed. The
explanation is as in the previous case: On the one hand, the DIFI
mode produces the oscillatory solution and, on the other hand, it
is responsible for the formation of FDS. Further increase of the
flow rate results in the stabilization of the FDS. Note that the
noise amplitude is very high: $\theta=1$. This is even higher then
the constant part $u_\inflow=0.1$ of the inlet perturbation. But
this can not prevent the stabilization. The pattern looks as in
Fig.~\ref{fSptmHopf}(c) and is not shown here.

As one can see from Fig.~\ref{fSptmDIFI}, the boundary layer of
the FDS manifests itself in a similar way as in the case shown in
Fig.~\ref{fVar}. The boundary layer effectively amplifies the
noise, so that the oscillatory solution appears near its right
edge. A structure without a boundary layer can also amplify noise,
but the area of amplification is very extended and grows when the
flow rate becomes higher. In particular, one can observe
spatio-temporal intermittency only far from the inlet. As in the
previous case, FDS without a boundary layer is stabilized at a
smaller flow rate in comparison to the structure that has a
boundary layer. The described behavior of FDS with and without a
boundary layer is typical. We have also analyzed the effect of
boundary layers for the two remaining cases and obtained similar
results.

\subsection{Case 3: Turing and Hopf modes with significant growth
rates}

\begin{figure}
  \centering
  \panelfig{a}{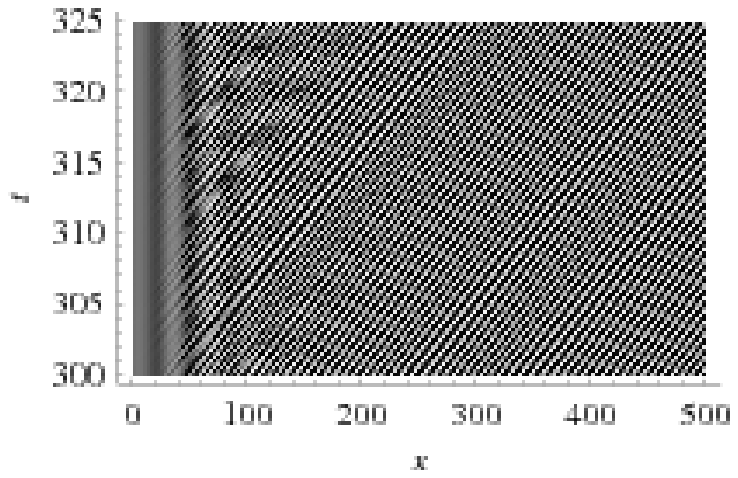}\\
  \panelfig{b}{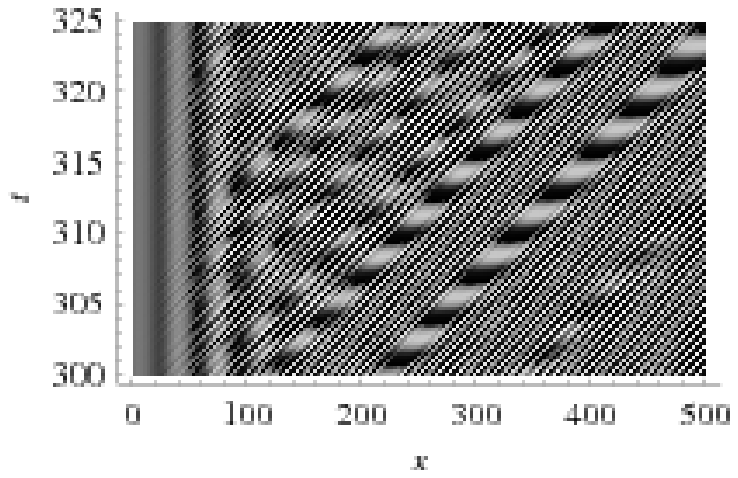}
  \caption{\label{fSptmTuring}FDS with noise in the presence of Turing
  and Hopf modes with significant growth rates.
  For all panels
  $a=20$, $b=5$, $\delta=20$, $\difadv=1$, $u_\inflow=0.5$, $\theta=0.1$
  ($b_\hpf=10.75$, $b_\trn\approx 21.928$).
  Panels correspond to different flow rates: a) $\phi=10$, b) $\phi=12$.}
\end{figure}

\begin{figure}
  \addtocounter{figure}{-1}
  \centering
  \panelfig{c}{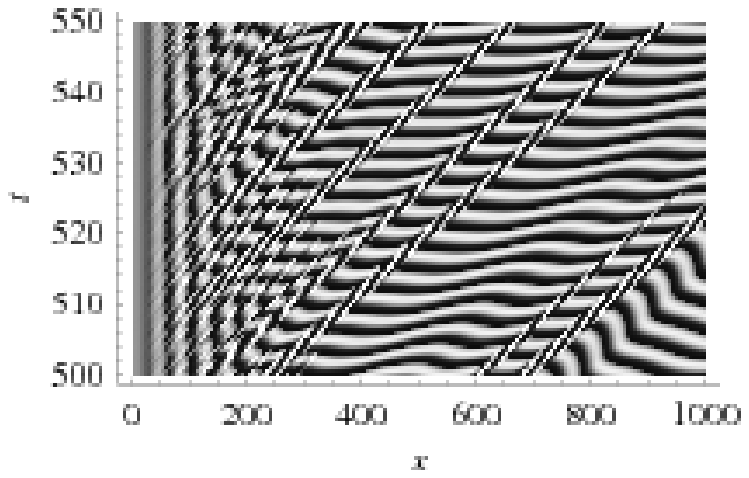}\\
  \panelfig{d}{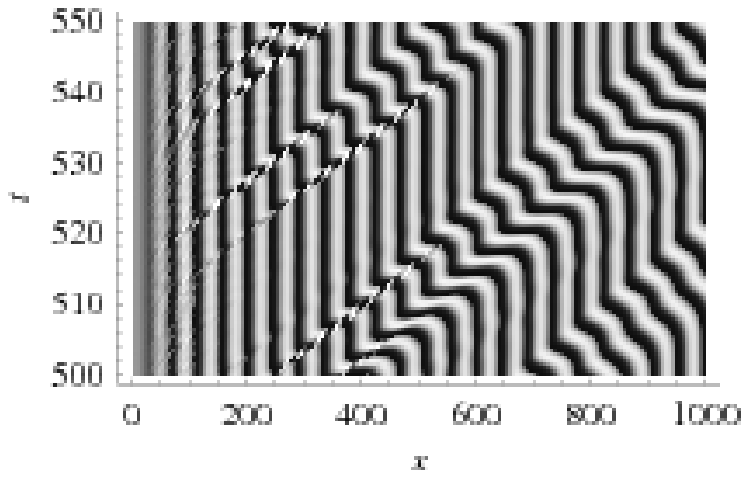}
  \caption{(\emph{continued}) The flow rates in the panels are c) $\phi=14$, d) $\phi=15.5$.}
\end{figure}

Now we assume that $b\ll b_\hpf < b_\trn$, so the system displays
both Hopf and Turing significant unstable modes and the Turing
mode has a higher growth rate. As follows from direct numerical
simulations, in a corresponding unbounded noiseless system the
Turing solution grows faster and suppresses the Hopf one.

Let us begin from small flow rates. When the FDS is affected by a
noise, the Turing mode suppresses both the FDS and Hopf solutions
and we observe almost perfect Turing structure behind a narrow FDS
area, see Fig~\ref{fSptmTuring}(a). This figure is qualitatively
similar to Figs.~\ref{fSptmHopf}(a) and \ref{fSptmDIFI}(a) where
unstable oscillatory modes also grow to form spatio-temporal
structures in place of a destroyed FDS.

Contrary to the previous cases, the higher flow rate in
Fig.~\ref{fSptmTuring}(b) does not result in the reestablishment
of FDS. It gives an advantage to the Hopf mode instead. This one
manifests itself as localized travelling areas of oscillations
embedded into the Turing structure.

Further increase of the flow rate makes the Hopf mode stronger,
see Fig.~\ref{fSptmTuring}(c). The Turing waves appear now as
localized perturbations to the Hopf structure. In the other words,
we have another spatio-temporal structure produced by the
strongest oscillatory mode that suppresses the FDS.

If the flow rate continues to grow, the Hopf solution starts to
give up its area to the FDS. This produces a spatio-temporal
intermittency, Fig.~\ref{fSptmTuring}(d). Note that a Turing
structures disappear in a different manner. In this figure there
are rare localized Turing waves that are swept by the flow and
decay in space. Sufficiently far from the inlet a combination of
oscillatory and stationary solutions appears in the same manner as
in the previous cases, see Figs.~\ref{fSptmHopf}(b) and
\ref{fSptmDIFI}(b). As the flow rate increases even further a
stabilization of the FDS takes place. We do not provide a
corresponding figure because the pattern looks like in
Fig.~\ref{fSptmHopf}(c).

\subsection{Case 4: Turing instability in presence of a weak Hopf mode}

\begin{figure}
  \centering
  \panelfig{a}{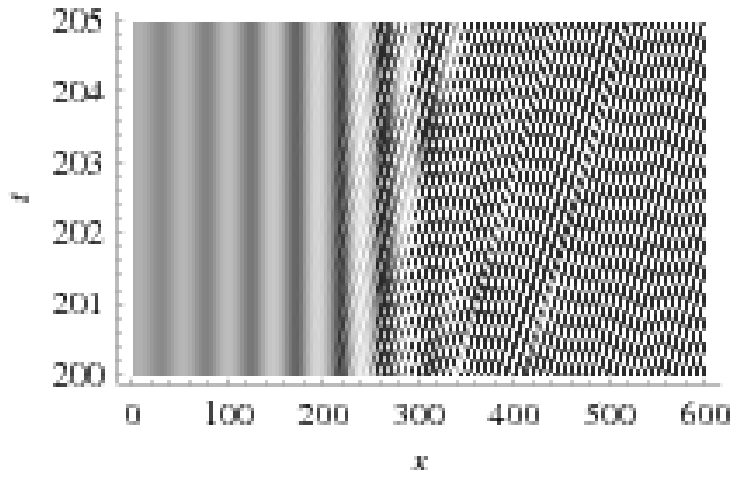}\\
  \panelfig{b}{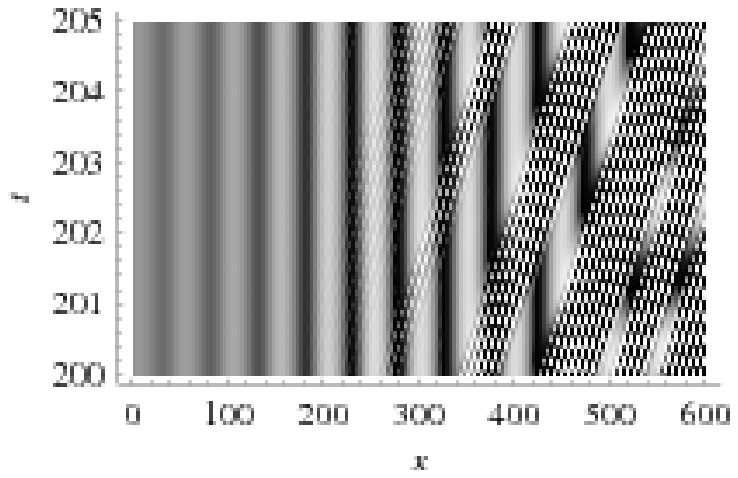}
  \caption{\label{fSptmTuring2}FDS with noise when a Turing mode is strong
  while a Hopf mode is weak.
  Observe the formation of
  Turing structure in place of the FDS in panel (a), $\phi=22$, and
  spatio-temporal intermittency in panel (b), $\phi=23.5$.
  The parameters are $a=25$, $b=10$ ($b_\hpf=14$, $b_\trn=37.549$), $\delta=25$, $\difadv=1$,
  $u_\inflow=0.5$, $\theta=10^{-5}$.}
\end{figure}

This case differs from three previous. Though a Hopf mode is
unstable and give rise to the formation of FDS, there is another
unstable mode, namely the Turing one, that has much higher growth
rate. When the flow rate is small, the spatio-temporal structure
is as in the previous case. Figure~\ref{fSptmTuring2}(a) shows how
a noise stimulates the growth of a Turing solution in place of the
FDS while the Hopf mode does not manifest itself at all (compare
with Fig.~\ref{fSptmTuring}(a)). But now even when the flow rate
becomes higher a Hopf solution does not emerge. The left end of
the Turing area becomes indented, and we observe an intermittent
structure, Fig.~\ref{fSptmTuring2}(b). Unlike three previous
structures, oscillatory and stationary fragments here have
different origins. The stationary FDS stripes appear due to the
Hopf mode while the oscillatory segments are produced by the
Turing mode. Hence, fragments of different solutions are not
joined into continuous, smooth zig-zag structures. They overlap
and partially suppress each other instead. When the flow rate
continues to grow, the FDS segments are enlarged while the Turing
areas becomes more and more rare. Finally, the FDS becomes stable
and occupies the whole space. The resulting pattern looks like in
Fig.~\ref{fSptmHopf}(c).

\section{\label{sTwoScen}Two scenarios of stabilization}

\begin{figure}
  \centering
  \panelfig{a}{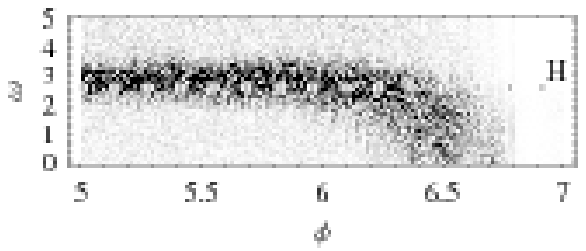} \\
  \panelfig{b}{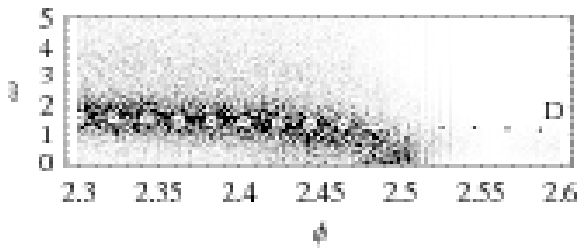} \\
  \panelfig{c}{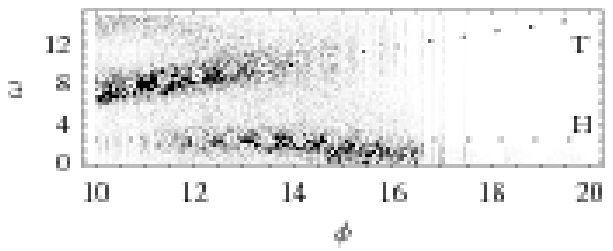} \\
  \panelfig{d}{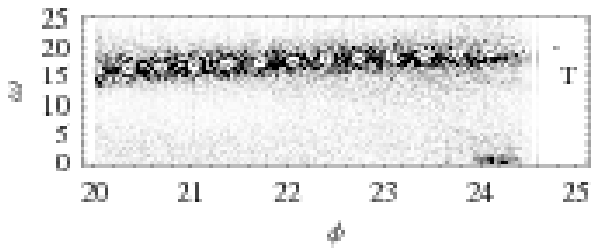}
  \caption{\label{fFour}Fourier spectra of temporal oscillations vs. flow rate at
  a fixed point near the outlet. Grey levels
  indicate values of Fourier amplitudes. Panels (a), (b), (c)
  and (d) (see below) correspond to Figs.~\ref{fSptmHopf}, \ref{fSptmDIFI},
  \ref{fSptmTuring} and \ref{fSptmTuring2},
  respectively. Dashed lines show Hopf (H), Turing (T) and
  DIFI (D) frequencies. Light areas on the right parts of the figures
  correspond to the stable FDS.}
\end{figure}

Four cases considered above demonstrate two different scenarios of
stabilization. The first one involves cases 1, 2 and 3, while the
second is represented by the case 4. A clearer evidence of this
can be provided with the Fourier analysis: We are going to
demonstrate the dependance of Fourier spectra of time oscillations
on the flow rate. This can be done with the following procedure. A
point is fixed near the outlet and a series of temporal
oscillations is recorded in this point. Then the mean value of the
series is subtracted and Fourier spectrum is computed using the
FFT algorithm. This is repeated for different flow rates and the
resulting Fourier amplitudes are plotted via grey scales in
Fig.~\ref{fFour}. The panels (a), (b), (c) and (d) in this figure
correspond to Figs.~\ref{fSptmHopf}, \ref{fSptmDIFI},
\ref{fSptmTuring} and \ref{fSptmTuring2}, respectively.

As we already discussed in our previous paper~\cite{KupSat05}, the
destruction of FDS takes place because this pattern can
selectively amplify a noise and the peak amplification corresponds
to the unstable oscillatory mode. This is confirmed in
Fig.~\ref{fFour}. We see here that Fourier components are gathered
around an unstable modes of the system. The resonant interaction
between selected harmonics and internal unstable modes results in
the destruction of FDS.

Figures~\ref{fFour}(a) and \ref{fFour}(b) demonstrate that the
stabilization in the Hopf and DIFI cases occurs due to the
detuning of the selected harmonics from the resonance. Bands of
selected frequencies are shifted down. Though the noise is still
amplified, but the selected frequencies differ from the resonant
and, hence, can not produce intensive excitation of the
oscillations. The amplitudes of the outlet oscillations are
reduced. We observe in the figure that the shaded areas becomes
lighter as the flow rate grows. This corresponds to an
intermittency in the spatio-temporal diagrams, see
Figs.~\ref{fSptmHopf}(b) and \ref{fSptmDIFI}(b). The shift of the
frequencies is reflected in these figures by the partial
straightening of the oscillatory segments that occur due to the
merging with vertical FDS stripes (compare, for example, the Hopf
structure in Fig.~\ref{fSptmHopf}(a) and the oscillatory areas in
Fig.~\ref{fSptmHopf}(b)).

Figure~\ref{fFour}(c) illustrates the same scenario of the
stabilization. This takes place via the detuning of selected
frequencies from the resonance with the Hopf mode. But here
another band of Fourier components appears around the Turing mode.
When the flow rate is small, noise is not amplified near the weak
Hopf mode. The resonance with the Turing mode is responsible for
the formation of the oscillatory solution in
Fig.~\ref{fSptmTuring}(a). For a higher flow rate both modes are
excited by the noise harmonics and we observe a spatio-temporal
structure as in Fig.~\ref{fSptmTuring}(b). Unlike the Hopf mode,
the selective amplification around the Turing frequency does not
depend directly on the flow rate and the Fourier components remain
gathered around the Turing mode. But the coefficient of
amplification vanishes as the flow rate grows: the shaded area in
the figure becomes lighter. The corresponding spatio-temporal
structure is shown in Fig.~\ref{fSptmTuring}(c). Finally, the
Turing band disappears and further stabilization proceeds as in
previous panels (a) and (b) of Fig.~\ref{fFour}. The corresponding
spatio-temporal diagram in Fig.~\ref{fSptmTuring}(d) is also
similar to the Hopf and DIFI cases, Figs.~\ref{fSptmHopf}(b)
and~\ref{fSptmDIFI}(b), respectively.

Fig.~\ref{fFour}(d) illustrates the second scenario of the
stabilization. The selective amplification takes place only around
the Turing mode, while the Hopf mode does not manifest itself
except the small area at $\phi\approx24$. The gathering of the
Fourier components near the Turing mode is similar to the case 3,
see Fig.~\ref{fFour}(c). But now this is the main mechanism
responsible for the stabilization. Observe that the shaded area in
Fig.~\ref{fFour}(d) disappears very sharply. This is explained by
an intermittent nature of analyzed time series, see
Fig.~\ref{fSptmTuring2}(b). When the flow rate approaches the
stabilization point, intervals between Turing bursts grow and
sooner or later we get an interval that is longer than the
observation time. In this point the characteristic Fourier
spectrum suddenly disappears. Though in the previous case 3 the
Turing oscillations also demonstrate a kind of intermittency with
the Hopf solution, see Fig.~\ref{fSptmTuring}(b), the
corresponding Fourier components in Fig.~\ref{fFour}(c) disappear
smoothly. It indicates that there the intervals between Turing
areas do not diverge.

\section{\label{sLin}Linear stability analysis}

In our previous paper~\cite{KupSat05} we have found that near the
critical point $\phi_\fdscrit$ an FDS pattern is linearly unstable
and any small perturbation applied at the inlet can destroy it. In
this section we study this effect and develop a linear stability
analysis of the FDS.

Let us first recall the discussions of Fig.~\ref{fVar}. This
figure displays spatial distributions of the variances of temporal
oscillations of $u$. The variances, both before and after the
stabilization point, grow exponentially within the boundary layer
of FDS. The stabilization reveals itself only for a fully
developed FDS while the boundary layer always amplifies noise. We
have observed this for many parameter values and believe that this
is typical for FDSs. Thus, we can neglect the boundary layer and
consider a fully developed periodic FDS solution
$u_\fdssol(x+P)=u_\fdssol(x)$ and $v_\fdssol(x+P)=v_\fdssol(x)$,
where $P$ is the period. To analyze the stability of this solution
we introduce a small perturbation, $u = u_\fdssol(x) + U(x)
e^{-i\omega t}$ and $v = v_\fdssol(x) + V(x) e^{-i\omega t}$,
where the frequency $\omega$ plays a role of an additional
parameter. $\omega$ is real because the perturbation appears due
to an external forcing that has a stationary amplitude. After the
substitution into Eqs.~(\ref{eLengEpst}), for small $U(x)$ and
$V(x)$ we obtain the following linear equations with periodic
coefficients:
\begin{subequations}
  \label{eFloqEq}
  \begin{eqnarray}
    U'' - \phi U' &=& ( 1-i\omega+4F_1(x) ) U + 4F_2(x) V,\\
    \delta V'' - \difadv \phi V' &=& b ( F_1(x)-1 ) U - (i\omega-b F_2(x) ) V,
  \end{eqnarray}
\end{subequations}
where $F_1=v_\fdssol(1-u_\fdssol^2)/(1+u_\fdssol^2)^2$ and
$F_2=u_\fdssol/(1+u_\fdssol^2)$. These second order equations have
complex coefficients. After straightforward transformations they
can be re-written as eight real equations of the first order.

Similar approach in a context of FDO/FDS patterns was developed by
McGraw and Menzinger~\cite{McGrawMenz05a} who studied small
perturbations to the steady state and considered linear modes
having real frequencies. This type of modes corresponds to a
stationary forcing at the inlet. The essential difference of our
study is the consideration of small perturbations to a \emph{fully
developed} FDS pattern. The linearized equations~(\ref{eFloqEq})
in this case have periodic coefficients and we should apply the
Floquet theorem to analyze its stability properties.

As follows from the Floquet theorem~\cite{JordSmith87} the system
has eight eigensolutions $\vec{\nu}(x)$ and eight corresponding
eigenvalues $\lambda$ for which
\begin{equation}\label{eFloqEig}
  \vec{\nu}(x+P)= \lambda \vec{\nu}(x).
\end{equation}
There are four couples of $\lambda$, either real and identical or
complex conjugated. Their absolute values can be labelled as
$|\lambda|_1 \leq |\lambda|_2 \leq |\lambda|_3 \leq |\lambda|_4$.
These values depend on the parameters of the system and on the
forcing frequency $\omega$. For $\phi>\phi_\fdscrit$ the two
highest values always correspond to diverging solutions of
Eqs~(\ref{eFloqEq}), $|\lambda|_3>1$ and $|\lambda|_4>1$, while
the others always represents a stable solution, $|\lambda|_1<1$.
The stability of FDS is characterized by $|\lambda|_2$. FDS is
stable if $|\lambda|_2 < 1$ for any $\omega$ while it is unstable
when $|\lambda|_2>1$ for certain values of $\omega$. It is more
convenient to consider a characteristic exponent that does not
depend on the period of FDS,
\begin{equation}
  \label{eCharExp}
  \rho_i = (\ln |\lambda|_i) / P.
\end{equation}
So, in a critical point of the linear stabilization the global
maximum of $\rho_2(\omega)$ passes zero.

We compute the eigenvalue $|\lambda|_2$ using the Floquet theorem.
The idea of this approach can be found, for example,
in~\cite{JordSmith87}. Let $\hat{A}$ be a linear evolution
operator advancing a solution over the period,
$\hat{A}\vec{\nu}(x)=\vec{\nu}(x+P)$. Each eigenvector $\vec{\nu}$
can be decomposed over the unit basis as $\vec{\nu}=\sum_{i=1}^8
\vec{e}_i c_i$, where $\vec{e}_i$ are the unit vectors and $c_i$
are the coefficients of the decomposition. Taking this into
account, we obtain from Eq.~(\ref{eFloqEig}):
\begin{equation}
  \sum_{i=1}^{8} \left( c_i \hat{A} \vec{e}_i \right) =
  \lambda \sum_{i=1}^{8} \left( c_i \vec{e}_i \right).
\end{equation}
These equations have a nontrivial solution for the coefficients
$c_i$ if $\lambda$ is an eigenvalue of the matrix with columns
$\hat{A}\vec{e}_i$. To obtain this matrix, we find numerically a
noiseless FDS solution and extract a full period. Then, we
interpolate it with polynomials and substitute the resulting
functions into Eqs.~(\ref{eFloqEq}). Starting from the unit
vectors $\vec{e}_i$ ($i=1\ldots8$), we find eight solutions to
these equations over the period and obtain the sought matrix.
Finally, we compute the eigenvalues and, after sorting by their
absolute values, take the second one.

\begin{figure}
  \centering
  \onefig{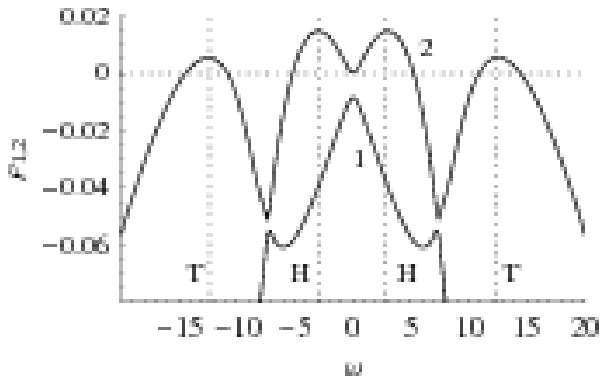}
  \caption{\label{fRho}Characteristic exponents $\rho_1$ and $\rho_2$ vs.
  $\omega$. Curve $\rho_2(\omega)$ has two positive maxima
  indicating the instability of corresponding FDS. Positions of the maxima coincide
  with the frequencies of Hopf and Turing modes that are indicated by the dashed lines
  ``H'' and ``T'', respectively.
  $a=15$, $b=5.5$, $\delta=10$, $\difadv=1$, $\phi=13$.}
\end{figure}

A numerical analysis shows that on the FDS onset $\phi =
\phi_\fdscrit$ the number of maxima of $\rho_2(\omega)$ coincides
with the number of linear modes, stable or unstable, of
Eqs.~(\ref{eLengEpst}). There is at least one couple of symmetric
maxima corresponding to Hopf or DIFI modes that are responsible
for the formation of FDS. Without a differential flow, the system
can also display a Turing mode. In this case $\rho_2(\omega)$
acquires another couple of maxima. For example in Fig.~\ref{fRho}
the first maxima from the origin correspond to the Hopf mode and
the second pair correspond to the Turing mode. All the maxima are
positive, hence the FDS is unstable.

The system~(\ref{eFloqEq}) can be considered as a linear amplifier
of an inlet forcing where $\rho_2(\omega)$ is an amplification
ratio. As follows from Fig.~\ref{fRho} the amplification is
selective. When a wide-band noise passes thorough the system,
noise harmonics grow if corresponding $\rho_2(\omega)$ is
positive. All other harmonics decay. As a result, oscillations at
the outlet consist of harmonics that are gathered around maxima of
$\rho_2(\omega)$, i.e. around linear modes of the initial
system~(\ref{eLengEpst}). This is remarkably coincides with our
qualitative observations in the above sections: when an FDS is
unstable inlet noise stimulates the growth of an unstable linear
mode to an oscillatory solution.

\begin{figure}
  \centering
  \onefig{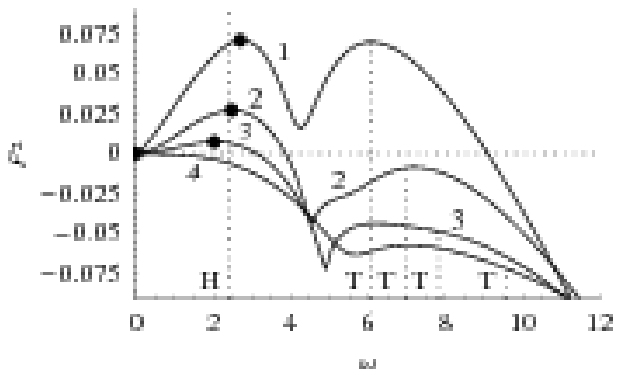}
  \caption{\label{fStab1}First scenario of the FDS stabilization.
  Bullet points show the successive positions of the right maximum
  (the left symmetric maximum is not shown) as the flow rate grows. Line ``H'' indicates
  the Hopf frequency and ``T'' mark the
  Turing frequencies for different flow rates.
  $a=15$, $b=4$, $\delta=10$, $\difadv=1$, $\phi_1=7$, $\phi_2=8$,
  $\phi_3=9$, $\phi_4=11$.}
\end{figure}

When the flow rate grows the structure of $\rho_2(\omega)$ varies
in two different ways. The first one is illustrated in
Fig.~\ref{fStab1}. This corresponds to the first scenario of the
FDS stabilization that is represented in Fig.~\ref{fFour}(a,b and
c) and has been discussed in Sec.~\ref{sTwoScen}. Different curves
in Fig.~\ref{fStab1} are plotted for different flow rates:

\begin{enumerate}
\item The curve $\rho_2$ has two positive maxima. Selected noise
harmonics excites oscillatory mode which destroy the FDS. Hence
the is linearly unstable.

\item The first maximum moves towards the origin while the
second one becomes negative. The Turing mode can not be excited
any more, but the other one (i.e., the Hopf mode in this figure,
but also it can be a DIFI mode) destroys the pattern. The system
is still unstable.

\item The first maximum continues the moving to origin. Thus,
frequencies of selected noise harmonics differ more and more from
the Hopf frequency. Because of this detuning the excitation of
Hopf oscillations becomes less effective.

\item Two symmetric maxima (the left one is not shown) merge
with a central minimum and these three extremums transform into a
single maximum at the origin. Now all the noise harmonics have
non-positive amplification ratio and those which are close to the
Hopf mode are even suppressed. As a result, the noise can not
excite oscillations. The FDS is linearly stable.
\end{enumerate}

Because the second couple of maxima passes zero first, it does not
play an important role in the stabilization. Thus, we classify as
the first scenario all cases when these maxima absent at all. For
example, these cases are shown in Fig.~\ref{fFour}(a,b).

\begin{figure}
  \centering
  \onefig{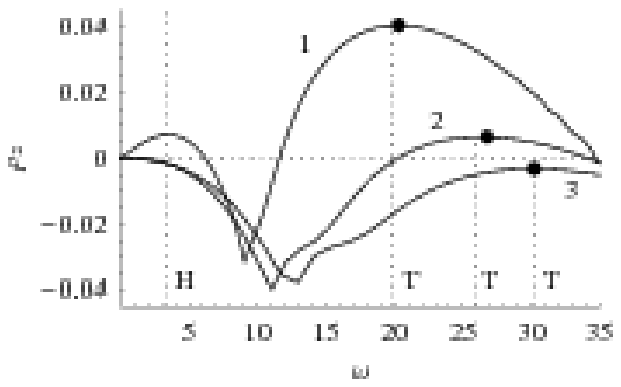}
  \caption{\label{fStab2}Second scenario of the linear stabilization of FDS.
  Bullet points show how the second maximum becomes negative as
  the flow rate grows.
  $a=20$, $b=9$, $\delta=20$, $\difadv=1$, $\phi_1=23$, $\phi_2=30$,
  $\phi_3=35$.}
\end{figure}

Fig.~\ref{fStab2} corresponds to the second scenario that has been
shown in Fig.~\ref{fFour}(d):

\begin{enumerate}
\item There are two positive maxima and the FDS is linearly unstable.

\item The first maxima merge (only the right part of the
symmetric curve is shown) but the second ones are still positive
and thus responsible for the instability of FDS.

\item The second couple of maxima pass
zero (only the right one is shown in the figure). As a result, no
oscillations are excited and the FDS becomes stable.
\end{enumerate}

So we see that two scenarios of the FDS stabilization that have
been revealed from numerical simulations now are confirmed by a
linear stability analysis.

\begin{figure}
  \centering
  \onefig{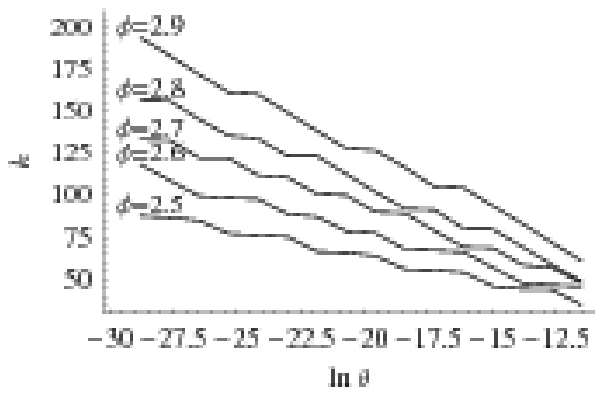}
  \caption{\label{fNoiseDist}Size of an area of a localized FDS $h$ vs.
  logarithm of the noise amplitude $\theta$.
  Observe linear dependance as predicts Eq.~(\ref{eNoiseDist})
  $a=20$, $b=3.5$, $\delta=3$, $\difadv=1$.}
\end{figure}

We can provide a direct verification of linear nature of the
instability of FDS near the critical point $\phi_\fdscrit$. Let us
suppose that the inlet perturbation is nearly equal to the
saturated FDS amplitude, so that we can neglect the boundary layer
of an FDS pattern. In the presence of inlet noise with an
amplitude $\theta$, the FDS occupies an area of a finite size, say
$h$. At the right edge of this area the noise reaches the
amplitude $\theta_1$ which is sufficient to destroy the FDS. If
the instability is linear, we can write $\theta_1=\theta e^{h
\rho_2}$, where $\rho_2\equiv \rho_2(\omega_{\mathrm{max}})$ is
the characteristic exponent responsible for the stability of FDS.
Hence, the size of an FDS area varies linearly with the logarithm
of the noise amplitude as
\begin{equation}\label{eNoiseDist}
  h=(\ln \theta_1 - \ln \theta)/ \rho_2.
\end{equation}
This equation can easily be verified directly. Solving numerically
Eqs.~(\ref{eLengEpst}) for different noise amplitudes, we can find
corresponding sizes of the FDS area. Fig.~\ref{fNoiseDist}
displays the resulting curves obtained for different flow rates.
As expected, $h$ decays linearly with $\ln \theta$.

Eq.~(\ref{eNoiseDist}) can be used for experimental measuring of
the critical flow rate of linear stabilization $\phi_\linstab$. A
factor in Eq.~(\ref{eNoiseDist}) is the reciprocal characteristic
exponent $\rho_2(\omega_{\mathrm{max}})$. To find this factor one
needs to vary amplitude of the inlet fluctuations and measure the
size $h$ of the FDS area. Unfortunately, it is not easy to perform
this practically. But because we deal with a linear problem, each
inlet harmonic is amplified separately. So we can substitute the
fluctuations with the most amplified harmonic\footnote{Experiments
where FDS is forced by a periodic oscillations at the inlet are
reported by K{\ae}rn and Menzinger~\cite{KernMenz00b,KernMenz02}
and also similar problem is considered by Taylor at
al.~\cite{TayBam02}.}. The size $h$ of the corresponding FDS area
should be equal to that producing by a wide-band inlet noise. The
frequency of this harmonic is $\omega_{\mathrm{max}}$. This value
may be calculated using the Floquet theorem as we have done above
for our system. Thus varying the amplitude $\theta$ of the inlet
forcing and measuring corresponding sizes $h$ one can obtain a
line $h(\ln \theta)$ and then calculate $\rho_2$ as an inverse
slope of this line. After that one needs to repeat the measuring
at different flow rates. Knowing $\rho_2$ vs. flow rate, one can
extrapolate the dependance $\rho_2(\phi)$ to a point where
$\rho_2$ vanishes and find the critical flow rate~$\phi_\linstab$.

\section{\label{sConcl}Summary and results}

We studied the stability of FDS patterns in presence of a noise at
the inlet. The following picture was revealed. If the flow rate is
not so high, fluctuations at the inlet stimulate the growth of an
oscillatory mode and thus prevents the formation of FDS. For
higher flow rates, however, an FDS pattern becomes linearly
stable. We found two scenarios of the linear stabilization: 1)
detuning of the most amplified noise harmonics from the resonance
with an oscillatory mode; 2) stopping of a noise amplification.

In our previous paper~\cite{KupSat05} we showed that linearly
stable FDS pattern still can be destructed by a noise if its
amplitude is sufficiently high. A threshold value of the noise
amplitude depends on the flow rate as a power law. Or vice versa:
for a fixed noise amplitude there is a flow rate above which an
FDS pattern becomes stable. This was referred to as a nonlinear
instability of FDS. In present paper we consider all possible
FDSs: 1) FDS is a stationary spatial distribution of Hopf
oscillations; 2) FDS emerges because of presence a DIFI; 3) FDS
based on the Hopf mode appears when also a Turing mode is present
and both Hopf and Turing modes have high growth rates; 4) FDS
appears when the Turing mode is much stronger then the Hopf mode.
These numerical experiments confirm our previous analysis. Also,
the numerical simulations reveal two scenarios of the
stabilization that remarkably correspond to those found within a
linear approach.

Let us consider an equation for a perturbation to FDS. (Now we
mean the nonlinear equation whose linearized version has been
studied in Sec.~\ref{sLin}, see Eq.~(\ref{eFloqEq}).) Our
observations can be summarized in terms of bifurcations of the
perturbation equation. A qualitative picture is following. When an
FDS is linearly unstable, the perturbation equation has a kind of
unstable fixed point. A bifurcation takes place when the flow rate
grows and our analysis predicts two scenarios for this. The result
of the bifurcation is the split of the unstable fixed point into a
stable fixed point and unstable limit cycle. This corresponds to a
nonlinear instability of the FDS. Small inlet noise decays because
the system remains close to the stable fixed point. But higher
noise brings out the system above the unstable limit cycle. An
amplification of noise results in the destruction of FDS. If we
increase the flow rate while keeping the noise amplitude
unaltered, a position of the limit cycle changes and the system
returns back to a vicinity of the stable fixed point. The unstable
limit cycle inherits properties of the unstable fixed point, so
that two scenarios of stabilization are observed when the system
passes this cycle. These are the scenarios that are obtained in
numerical experiments in Secs.~\ref{sStab} and \ref{sTwoScen}.

The destruction of FDS in presence of an inlet noise takes place
because this periodic pattern serves as a selective amplifier of
the forcing. The highes amplification takes places at frequency
equal to an unstable oscillatory mode. This resonant interaction
excites the oscillatory solution that suppresses the FDS. Hence,
the specific type of an inlet forcing is not very important. The
sufficient requirement is that its spectrum contain harmonics
close the resonance. In particular, it means that our results
remains valid for a Gaussian noise. We tried this type of noise.
The results were qualitatively similar.

A pure periodic forcing at the inlet is already studied well. As
reported in~\cite{KernMenz00b,KernMenz02,TayBam02}, the inlet
oscillations of concentrations stimulate the destruction of FDS
provided that the frequency is close to the natural frequency of a
system. In the other words, the destruction takes place when the
forcing is in the resonance with a mode responsible for the
formation of FDS. A minor decreasing of the inlet frequency from
the resonance results in the pulsating waves as
in~\cite{KernMenz00b} or in the zigzag patterns~\cite{TayBam02}.
These structures are very similar to our intermediate situations
represented in Figs.~\ref{fSptmHopf}(b) and \ref{fSptmDIFI}(b).
These results, in fact, corresponds to our first scenario of the
FDS stabilization via detuning from the resonance.

Also the resonance with the natural frequency of a system is
reported by McGraw and Menzinger~\cite{McGrawMenz05b}. They apply
a periodic modulation to the flow rate and observe a localized FDS
area suppressed downstream by an oscillatory solution. It is
interesting that the suppression takes place not only at the
natural frequency. The resonance and further suppression is also
observed at nonlinear rations $2:1$ and $3:1$!

Finally, we suggest an idea for an experimental measuring of a
flow rate of linear stabilization. Applying to the FDS inlet a
weak periodic forcing with specific frequency one can find a
characteristic exponent that is responsible for the linear
instability. One needs to find this exponent for different flow
rates and then extrapolate the dependance to a point where the
exponent vanishes. A corresponding flow rate is the sought
critical value of the linear stabilization.

This work was partially supported by grant of RFBR (no.
04-02-04011).

\bibliographystyle{elsart-num}
\bibliography{fds_noise}

\begin{thebibliography}{10}
\expandafter\ifx\csname url\endcsname\relax
  \def\url#1{\texttt{#1}}\fi
\expandafter\ifx\csname urlprefix\endcsname\relax\def\urlprefix{URL }\fi

\bibitem{KuzMos97}
S.~P. Kuznetsov, E.~Mosekilde, G.~Dewel, P.~Borckmans, Absolute and convective
  instabilities in a one-dimensional {B}russelator flow model, J. Chem. Phys.
  106 (1997) 7609--7616.

\bibitem{AndrBache99}
P.~Andres{\'e}n, M.~Bache, E.~Mosekilde, G.~Dewel, P.~Borckmans, Stationary
  space periodic structures with equal diffusion coefficients, Phys. Rev. E 60
  (1999) 297--301.

\bibitem{KernMenz99}
M.~K{\ae}rn, M.~Menzinger, Flow-distributed oscillations: {S}tationary chemical
  waves in a reacting flow, Phys. Rev. E 60 (1999) R3471--R3474.

\bibitem{KernMenz00b}
M.~K{\ae}rn, M.~Menzinger, Pulsating wave propagation in reactive flows:
  {F}low-distributed oscillations, Phys. Rev. E 61 (2000) 3334--3338.

\bibitem{KernMenz02}
M.~K{\ae}rn, M.~Menzinger, Experiments on flow-distributed oscillations in the
  {B}elousov--{Z}habotinsky reaction, J. Phys. Chem. A 106~(19) (2002)
  4897--4903.

\bibitem{SatMaini01}
R.~A. Satnoianu, P.~K. Maini, M.~Menzinger, Parameter space analysis, pattern
  sensitivity and model comparison for {T}uring and stationary flow-distributed
  waves ({FDS}), Physica~D 160 (2001) 79--102.

\bibitem{Sat03}
R.~A. Satnoianu, Coexistance of statonary and traveling waves in
  reaction--diffusion--advction systems, Phys. Rev. E 68 (2003) 032101.

\bibitem{McGrawMenz05a}
P.~N. McGraw, M.~Menzinger, Pattern formation by boundary forcing in
  convectively unstable, oscillatory media with and without differential
  transport, Phys. Rev. E 72 (2005) 026210.

\bibitem{SatMenz00}
R.~A. Satnoianu, M.~Menzinger, Non-{T}uring stationary patterns in
  flow-distributed oscillators with general diffusion and flow rates, Phys.
  Rev. E 62 (2000) 113--119.

\bibitem{Kup04}
P.~V. Kuptsov, Rigid transition to the stationary structure and imposed
  convective instability in a reaction--diffusion system with flow, Physica D
  197 (2004) 174--195.

\bibitem{MigSat06}
D.~G. M{\'\i}guez, R.~A. Satnoianu, A.~P. Mu{\~n}uzuri, Experimental steady
  pattern formation in reaction-diffusion-advection systems, Phys. Rev. E
  73~(2) (2006) 025201.

\bibitem{MigIzus06}
D.~G. M{\'\i}guez, G.~G. Iz{\'u}s, A.~P. Mu{\~n}uzuri, Robustness and stability
  of flow-and-diffusion structures, Phys. Rev. E 73~(1) (2006) 016207.

\bibitem{KupKuz02}
P.~V. Kuptsov, S.~P. Kuznetsov, E.~Mosekilde, Particle in the brusselator model
  with flow, Physica~D 163 (2002) 80--88.

\bibitem{BamToth02}
J.~R. Bamforth, R.~Toth, V.~Gaspar, S.~K. Scott, Scaling and dynamics of ``flow
  distributed oscillation patterns'' in the {B}elousov--{Z}habotinsky reaction,
  PCCP 4~(8) (2002) 1299--1306.

\bibitem{McGrawMenz05b}
P.~N. McGraw, M.~Menzinger, Flow-distributed oscillation, flow-velocity
  modulation, and resonance, Phys. Rev. E 72 (2005) 027202.

\bibitem{KupSatDan05}
P.~V. Kuptsov, R.~A. Satnoianu, P.~G. Daniels, Pattern formation in 2{D}
  reaction--diffusion channel with {P}oiseuille flow, Phys. Rev. E 72 (2005)
  036216.

\bibitem{TayBam02}
A.~F. Taylor, J.~R. Bamforth, P.~Bardsley, Complex pattern development in a
  plug-flow reactor, PCCP 4~(22) (2002) 5640--5643.

\bibitem{Saar88}
W.~van Saarloos, Front propagation into unstable states: {M}arginal stability
  as a dynamical mechanism for velocity selection, Phys. Rev. A 37 (1988)
  211--229.

\bibitem{Saar89}
W.~van Saarloos, Front propagation into unstable states. {II}. {L}inear versus
  nonlinear marginal stability and rate of convergence, Phys. Rev. A 39 (1989)
  6367--6390.

\bibitem{KupKuz03}
P.~V. Kuptsov, S.~P. Kuznetsov, C.~Knudsen, E.~Mosekilde, Absolute and
  convective instabilities in the one-dimensional {B}russelator model with
  flow, Recent Res. Devel. Chem. Physics 4 (2003) 633--658.

\bibitem{Deissler85}
R.~J. Deissler, Noise-sustained structure, intermittency, and the
  ginzburg–landau equation, J. Stat. Phys. 40 (1985) 371--–395.

\bibitem{DeissFarm92}
R.~J. Deissler, J.~D. Farmer, Deterministic noise amplifiers, Physica D 55
  (1992) 155--–165.

\bibitem{BorckDew95}
P.~Borckmans, G.~Dewel, A.~D. Witt, D.~Walgraef, The differential flow
  instabilities, in: K.~S. R.~Kapral (Ed.), Chemical Waves and Patterns, Kluwer
  Academic Publishers, Dordrecht, 1995, pp. 87--95.

\bibitem{Landa96}
P.~S. Landa, Turbulence in nonclosed fluid flows as a noise-induced phase
  transition, Europhys. Lett. 36~(6) (1996) 401--406.

\bibitem{Kuz02}
S.~P. Kuznetsov, Noise--induced absolute instability, Mathematics and Computers
  in Simulation 58 (2002) 435–--442.

\bibitem{KupSat05}
P.~V. Kuptsov, R.~A. Satnoianu, Stability of {F}low--~and {D}iffusion
  {D}istributed {S}tructures ({F}{D}{S}) to inlet noise effects, Phys. Rev. E
  71 (2005) 015204.

\bibitem{LengEpst91}
I.~Lengyel, I.~Epstein, Modelling of {T}uring structures in the
  chlorite--iodide--malonic acid--starch reaction system, Science 251 (1991)
  650--Ö652.

\bibitem{JenPanMos94}
O.~Jensen, V.~O. Pannbacker, E.~Mosekilde, G.~Dewel, P.~Borckmans, Localized
  structures and front propagation in the {L}engyel--{E}pstein model, Phys.
  Rev. E 50~(2) (1994) 736--749.

\bibitem{Turing52}
A.~M. Turing, The chemical basis of morphogenesis, Philos. Trans. R. Soc. B 237
  (1952) 37--72.

\bibitem{NicPrig77}
G.~Nicolis, I.~Prigogine, Self-Organization in Nonequilibrium Systems, Wiley,
  New York, 1977.

\bibitem{KernMenzSat02}
M.~K{\ae}rn, M.~Menzinger, R.~A. Satnoianu, A.~Hunding, Chemical waves in open
  flows of active media: {T}heir relevance to axial segmentation in biology, J.
  R. Chem. Soc., Faraday Discussions 120 (2002) 295--312.

\bibitem{RovMenz93}
A.~B. Rovinsky, M.~Menzinger, Self--organization induced by the
  differential--flow of activator and inhibitor, Phys. Rev. Lett. 70 (1993)
  778--781.

\bibitem{AndrMos00}
P.~Andres{\'e}n, E.~Mosekilde, G.~Dewel, P.~Borckmans, Comment on
  ``{F}low-distributed oscillations: {S}tationary chemical waves in a reacting
  flow'', Phys. Rev. E 62~(2) (2000) 2992--2993.

\bibitem{KernMenz00}
M.~K{\ae}rn, M.~Menzinger, Reply to ``{C}omment on `{F}low-distributed
  oscillations: {S}tationary chemical waves in a reacting flow' '', Phys. Rev.
  E 62~(2) (2000) 2994--2995.

\bibitem{JordSmith87}
D.~W. Jordan, P.~S. Smith, Nonlinear ordinary differential equations, Clarendon
  press, Oxford, 1987.

\end{thebibliography}

\end{document}